# Sequence-Defined Phase Behavior of Poly(N-Isopropylacrylamide-co-Acrylamide) in Water

Sandeep Parma,[1,2] R. Nagaranajan[1] and Tarak K Patra[1,3*]

[1]Deparment of Chemical Engineering, Indian Institute of Technology Madras, Chennai, TN 600036, India
[2]Department of Chemical Engineering, Manipal Institute of Technology, Manipal Academy of Higher Education, Manipal, KA 576104, India
[3]Center for Atomistic Modelling and Materials Design, Indian Institute of Technology Madras Chennai, TN 600036, India

**Abstract**

The precise arrangement of different chemical moieties in a polymer determines its thermophysical properties. How the sequence of moieties impacts the properties of a polymer is an outstanding problem in polymer science. Herein, we address this problem for the thermoresponsive property of poly(N-isopropylacrylamide-co-acrylamide) in water using all-atom molecular dynamics (MD) simulations. Eight distinct copolymers, each with a different arrangement of NIPAM(N-isopropylacrylamide) and AM (acrylamide) monomers, are considered. The lower critical solution temperature (LCST) shows a strong correlation with the mean block length of the periodic sequences of the copolymer. We further identify copolymer sequences that exhibit both the LCST and upper critical solution temperature (UCST). Moreover, there are sequences that do not show any LCST or UCST for the temperature range explored in this study. This wide variability in thermorepsonsive property is found to be closely linked to the the extent of hydrongen bond formation in the system, which appears to have a significant correlation with the monomer sequence of the copolymer. These findings offer new directions in the design of structurally diverse thermoresponsive copolymers.

*Corresponding Author, E-mail: tpatra@iitm.ac.in

# I. Introduction

A defining characteristic of thermoresponsive polymers is their ability to undergo solubility transitions in response to temperature changes. Polymers that exhibit increased solubility with rising temperature are classified as exhibiting upper critical solution temperature (UCST) behavior. This behaviour characterizes their capacity to dissolve more readily in a solvent above a specific temperature threshold. Conversely, some polymers demonstrate a decrease in solubility as the temperature increases. This phenomenon, widely recognised as the lower critical solution temperature (LCST) behavior, reflects the point at which the polymer becomes less soluble, leading to phase separation or precipitation. Both UCST and LCST behaviors are critical for applications in smart materials, drug delivery systems, and temperature-sensitive devices. A common example of thermoresponse polymer is Poly(N-isopropylacrylamide) (PNIPAM)[1–5], which is water-soluble at low temperature due to the polymer-water hydrogen bonding. It becomes hydrophobic as water-polymer hydrogen bonds break with an increase in temperature, and polymer-polymer interactions dominate, leading to polymer collapse in water. Thus, below the LCST point, PNIPAM chains are in a hydrated state with an expanded structure (coil)[6], and above the LCST point, it becomes a globule conformation. This phase transition is reversible, and the critical temperature can be modulated by a variety of techniques. The LCST of PNIPAM appears to have a strong correlation with its molecular weight.[7] The backbone and side chain chemistry have been shown to alter their LCST behaviour.[8] The free radical copolymerization of NIPAM with various imidazolium-based ionic liquids provides a wide range of LCST behaviour.[6] The presence of AM moieties in the backbone of PNIPAM is found to increase the LCST.[9] In contrast, the presence of methylthioethyl acrylate suppresses the LCST of PNIPAM.[10] The block length of a PNIPAM-$b$-poly ($N$-vinylcaprolactam) can tune its LCST point.[11] Poly[oligo(ethylene oxide)] based gradient and random copolymers exhibit distinct LCST behaviours.[12] These studies clearly demonstrate that introducing different chemical groups into the topology of a PNIPAM chain can significantly influence its phase behaviour. However, while the overall composition has been studied, the role of the relative positions of these chemical groups along the polymer backbone—that is, the precise sequence of the copolymer—remains significantly underexplored. A deeper understanding of this sequence-level control is crucial for advancing rational design of functional polymer materials.

In general, sequence-defined synthetic polymers represent an emerging area of research that offers tremendous opportunities for the design of novel materials with precisely tailored properties. Recent advancements in synthetic chemistry have made it possible to achieve an unprecedented level of control over the sequence of different chemical moieties within a copolymer molecule.[13–16] These precision polymers offer vast opportunities for molecular-scale material design, providing diverse microstructures, phase behaviours, functionalities, and activities in polymers. In our recent studies, we have shown that the three-dimensional structure of a copolymer can be extensively controlled by fine-tuning its monomer sequence.[17,18] Consequently, the bulk properties of polymeric materials—such as conductivity, rigidity, biodegradability, antibacterial activity, elasticity, and optoelectronic characteristics—can be precisely tailored by modifying the monomer sequence in their constituent polymers. For instance, copolymers with different sequences are used to enhance the thermodynamic stability of polymer mixtures, making them highly applicable in emulsions and composite materials.[19–22] Likewise, the arrangement of monomers in a charged copolymer influences its structural phase transitions,[23,24] the liquid–liquid phase transition and its corresponding critical points.[25] With these advancements in the field, here we investigate to what extent the LCST of PNIPAM can be modulated by doping AM into its structure.

We address the sequence-specificity of the NIPAM-AM copolymer solubility in water using all-atom MD simulations. All-atom MD simulations have been extensively used to understand the thermodynamic driving forces for the phase transition of PNIPAM and other polymers.[26–31] Building upon these prior works, here we consider eight different types of NIPAM-AM copolymer sequences with a fixed compostion along with the pure PNIPMA and PAM. All copolymer chains considered in this study are composed of 30 monomer units. The copolymers are made using an equal ratio of NIPAM and AM, with each chain containing 15 NIPAM and 15 AM monomers. Given this composition and chain length, only four distinct periodic arrangements are possible. We systematically include all of these periodic copolymers in our analysis. In contrast, the number of non-periodic sequences that can be generated with this composition and chain length is exceedingly large, making comprehensive sampling impractical. To capture a representative diversity in non-periodic sequence architecture, we select four such copolymer sequences: one gradient copolymer, one random copolymer, and two triblock copolymers—one with AM blocks at both termini and NIPAM in the center, and the other with the reverse arrangement (NIPAM ends and AM center). We study their phase behaviour in water for the temperature range of *T=260 K* to *T=360 K*. We find that the PNIPAM and PAM exhibit LCST and UCST phase behaviour, respectively, within this temperature range, which is consistent with the literature. However, we observe a multitude of phase behavior among these eight copolymers, including pure LCST, a combination of LCST and UCST, and no-phase change within the temperature range. We correlate this diversity of pahse behaivor with the variability of the hydrogen bond network with sequences. Our work demonstrates how the molecular interaction can be tuned with the copolymer sequence, and its implications in the phase behavior of the NIPMA-AM copolymer.

## II.   Model and Methodology

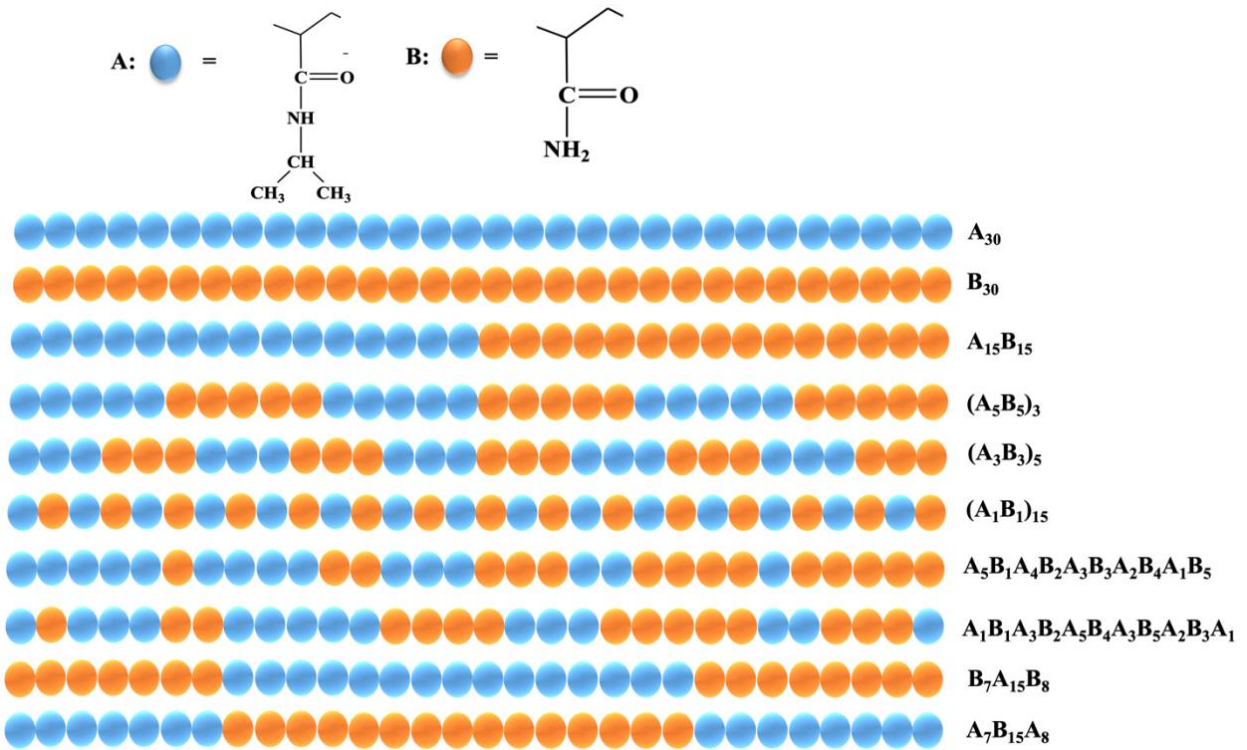

*Figure 1:Two chemcial moieties- NIPMA and AM are shemcatically shown at th top. The NIPAM and AM correspond to A and B-type moieties, respectively. They are represented as blue and orange beads, respectively. All the seqeucnes which are considered in this study, are shwn in the bottm. All polymers are of chain lenth 30, comprising 15 A groups and 15 B groups.*

Schematic representations of all the polymers are shown in Figure 1. All the polymers are atactic in our study. We build the topologies of polymers using CHARMM[32] and CHARMM-GUI[33,34] packages. We solvate these structures in cubic simulation boxes with a side length of 10 nm with TIP3P[35] water molecules. A simulation box consists of about 24700 atoms, and it is periodic in all three directions. In these simulations, we use a single polymer chain in water. The NIPAM and AM are modelled using the CHARMM force field[36]. We use the particle mesh Ewald method to calculate the long-range electrostatic interactions in the system. Initially, the energy in each system is minimised using the steepest descent algorithm with 10,000 steps to remove any overlap of atoms. In the next stage, the system is equilibrated in an isothermal-isobaric ensemble (NPT) for 40 ns followed by a production run of 10 ns. The system temperature is controlled using the Bussi−Donadio−Parrinello velocity-rescaling thermostat[37], which is a modified Berendsen thermostat[38] containing an additional stochastic term that allows for the correct sampling of the kinetic energy distribution, with a time constant of 0.1 ps. The system pressure is maintained isotropically at 1 bar using the C-rescale barostat [39] with a time constant of 2.0 ps. The integration time step is 2 fs. Each system is simulated for 12 different temperatures in the range of 260K to 360K. All the simulations are done within the MD environment of the GROMACS package.[40] All the properties are averaged over three independent simulations.

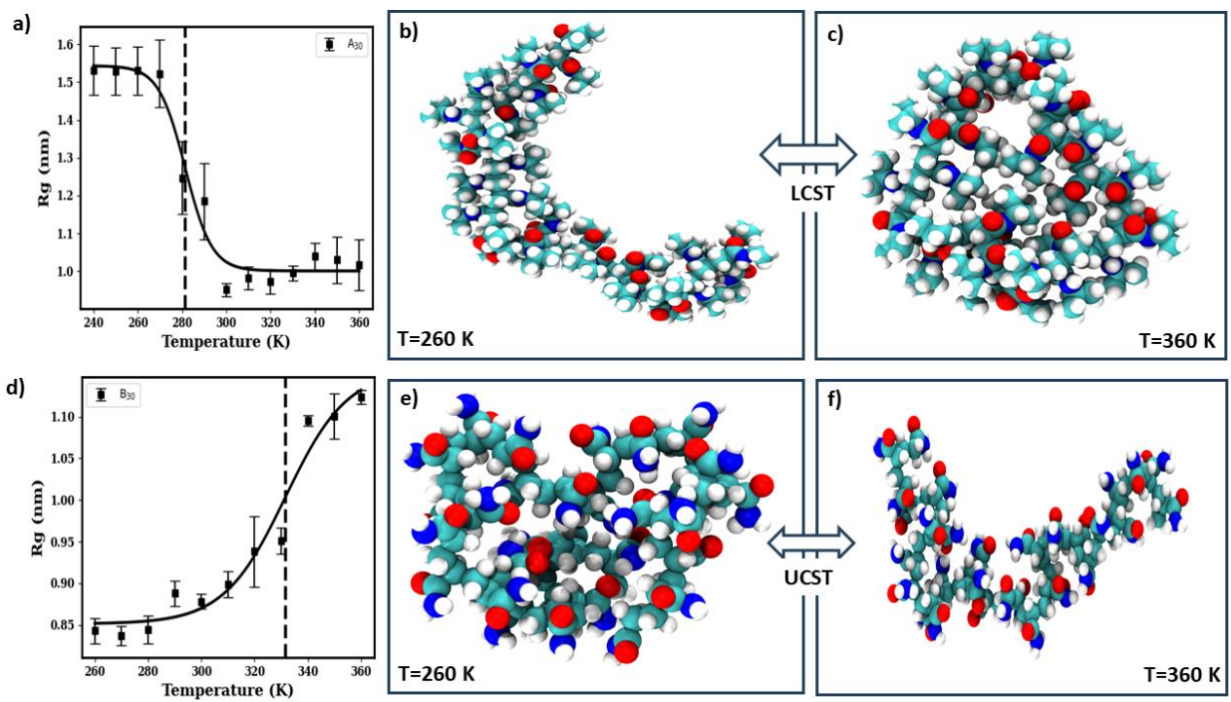

*Figure 2: The $R_g$ is plotted as a function of temperature for the PNIPAM-water system in (a). Two MD snapshots of the PNIPAM are shown in (b) and (c) for T=260 K and T=360 K, respectively. The $R_g$ as a function of temperature is plotted in (d) for PAM - water system. The MD snapshots of PAM at T=260K and T=360K are shown in ( e) and (f), respectively. Water molecules are deleted for the MD snapshots for visual clarity.*

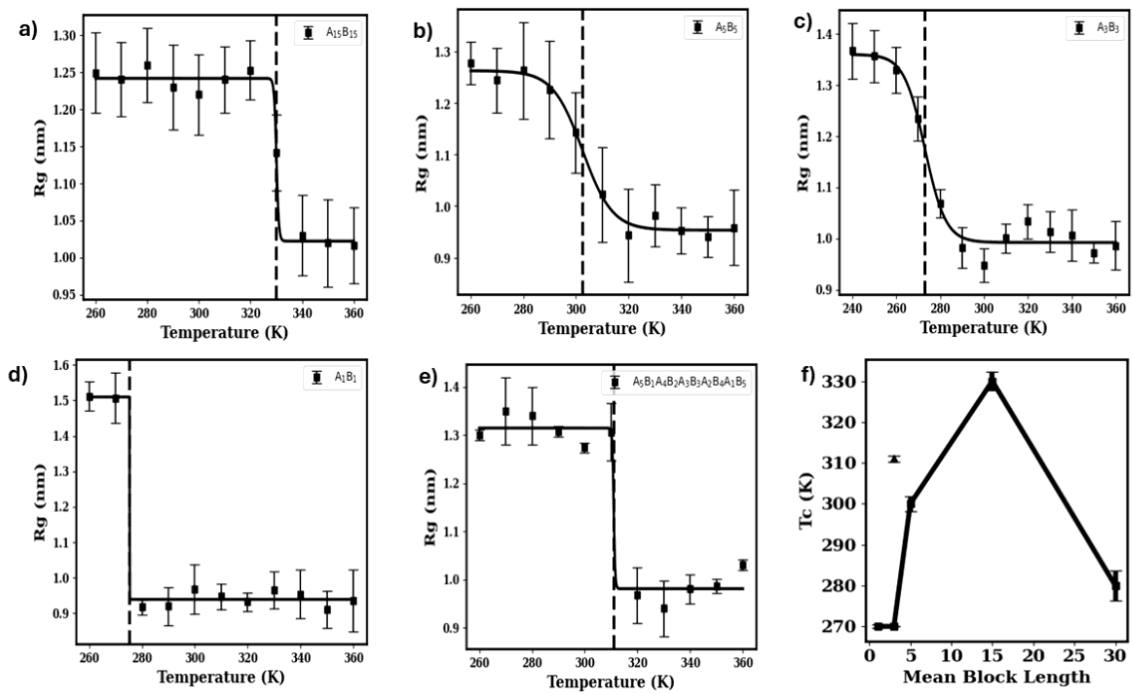

Figure 3: The radius of gyration ($R_g$) is ploted as a function of temperature in (a-e) for five copolymers as metioned in the panels. The LCST ($T_c$) as function of mean block length of a copolymer is plotted in (f).

## III. Results and Discussion

The radius of gyration ($R_g$) is the key structural descriptor used to evaluate the conformational behaviour of polymer chains in water. The $R_g$ of the PNIPAM and PAM in water over a temperature range from 260 K to 360 K is shown in Figure 2 along with representative MD snapshots. For the PNIPAM, the Rg decreases monotonically with increasing temperature, indicating a coil-to-globule transition of the system. In contrast, PAM showed an opposite trend in Rg over the same temperature range, as its hydrophilicity increases with the increase of temperature, leading to a globule-to-coil transition. We estimate the critical points of these transitions by fitting the data to a sigmodal function of the form $R_g(T) = \frac{a-b}{1+exp\left(\frac{T-T_c}{\Delta T}\right)} + b$. Here, $a$ and $b$ correspond to the $R_g$ at the lowest and highest temperatures, respectively. The $T_c$ and $\Delta T$ are the critical temperature and slope factor of the crossover. The transition from the coil to globule takes place at $T_c$=280K, which is the LCST of the PNIPAM system. For the PAM, the transition occurs at $T_c$=330 K, which is considered as the UCST of the system. Next, we present the $R_g$ as a function of temperature for six copolymers in Figure 3. All of them show an LCST-type phase behaviour wherein the $R_g$ decreases abruptly with the temperature of the system. We estimate the critical temperatures of the transition for all of these cases by fitting the sigmoidal curve, similar to the homopolymer cases. We plot the LCST as a function of mean block length (MBL) in Figure 3f. The MBL is defined as the average size of all the *A* and *B* blocks in a chain. The observed variation in LCST with MBL reveals a non-monotonic trend. It appears that the $T_c$ is correlated with MBL for the set of periodic sequences. As the MBL increases, the LCST initially rises and reaches a maximum at MBL = 15. This increase may be attributed to the enhanced hydrophilicity or stronger intramolecular interactions that delay the collapse transition. However, beyond this composition, when the MBL becomes 30, which is a pure PINIPAM (homopolymer), the LCST drops significantly. This is due to the changes in chain conformation, which corresponds to intermolecular interactions that favour aggregation at a lower temperature. This behaviour highlights the complex interplay between copolymer sequence and phase behaviour. It suggests that the LCST of a cpolymer involves more nuanced molecular interactions, which can be tuned with its sequences. Interestingly, the $T_c$ corresponding to the non-periodic sequence - $A_5B_1A_4B_2A_3B_3A_2B_4A_1B_5$ deviates significantly from this apparent parabolic trend observed for the periodic sequences, as shown in Figure 3f. This outlier behaviour underscores the limitation of using mean block length as a sole descriptor for predicting phase behaviour in sequence-defined copolymers. While mean block length provides a coarse measure of sequence-specificity, it fails to capture the underlying complexity introduced by the precise monomer arrangement. These results suggest that more sophisticated sequence metrics—such as block distribution, sequence asymmetry, or local composition fluctuations—may be necessary to more accurately correlate sequence architecture with thermal responsiveness. To better understand the phenomenon, we perform a detailed analysis of hydrogen bond formation in the system during the phase transition. We consider oxygen atoms of the polymer to be hydrogen bond acceptors, and nitrogen atoms of the polymer and oxygen atoms of water molecules to be able to accept and donate hydrogen bonds. We adopt geometric criteria of an acceptor-donor distance to determine a hydrogen bond. A hydrogen bond is considered to be formed when the distance between the donor and acceptor atoms is $r_{HB} \leq 0.35$ nm, and the angle formed by the acceptor, donor, and hydrogen atoms is $\alpha \leq 30°$. The average numbers of hydrogen bonds between the polymer and water molecules are plotted in Figure 4. This indicates that the number of polymer-water hydrogen bonds decreases monotonically with increasing temperature for all the cases. As the temperature rises, the hydrogen-bonding network connecting monomers and water molecules progressively breaks down, ultimately leading to the collapse of the polymer chain.

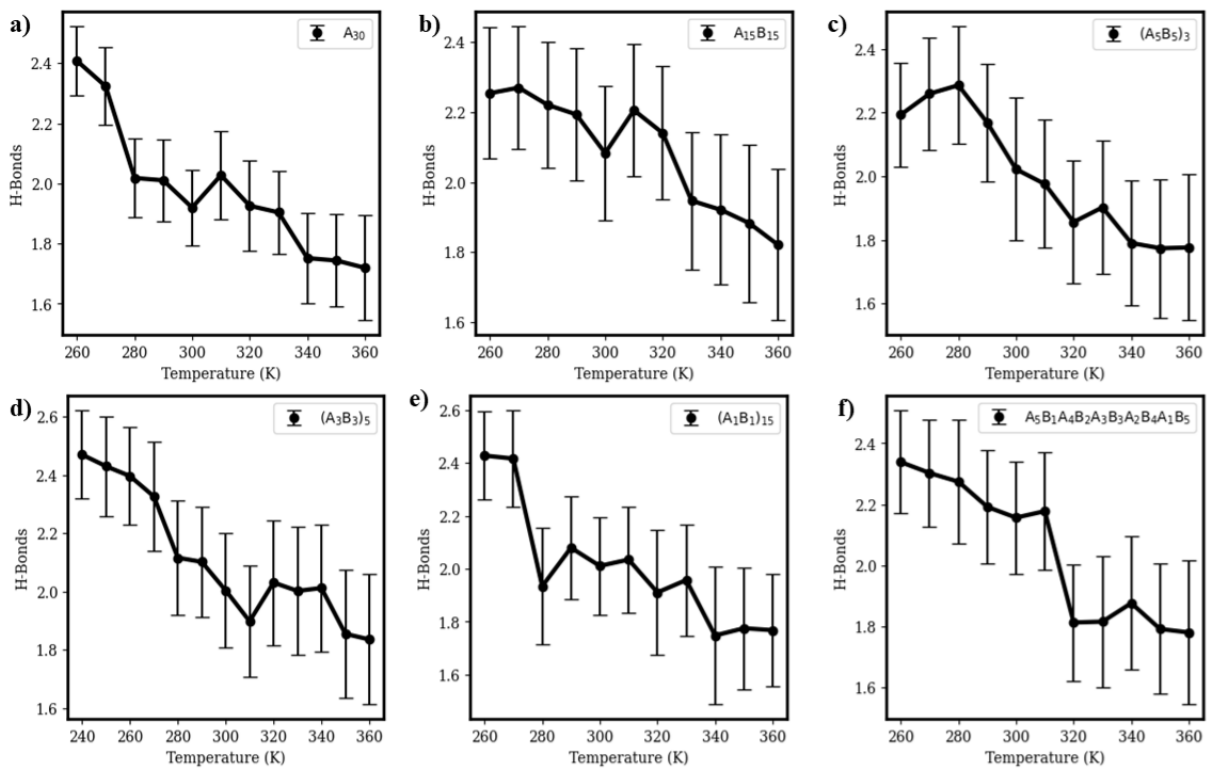

*Figure 4: The number of polymer-water hydrogen bonds in the system is plotted as a function of temperature for 6 polymers as mentioned in the panels for temperature ranging from 260K to 360K. The number of hydrogen bonds is normalised by the number of monomers in a polymer chain.*

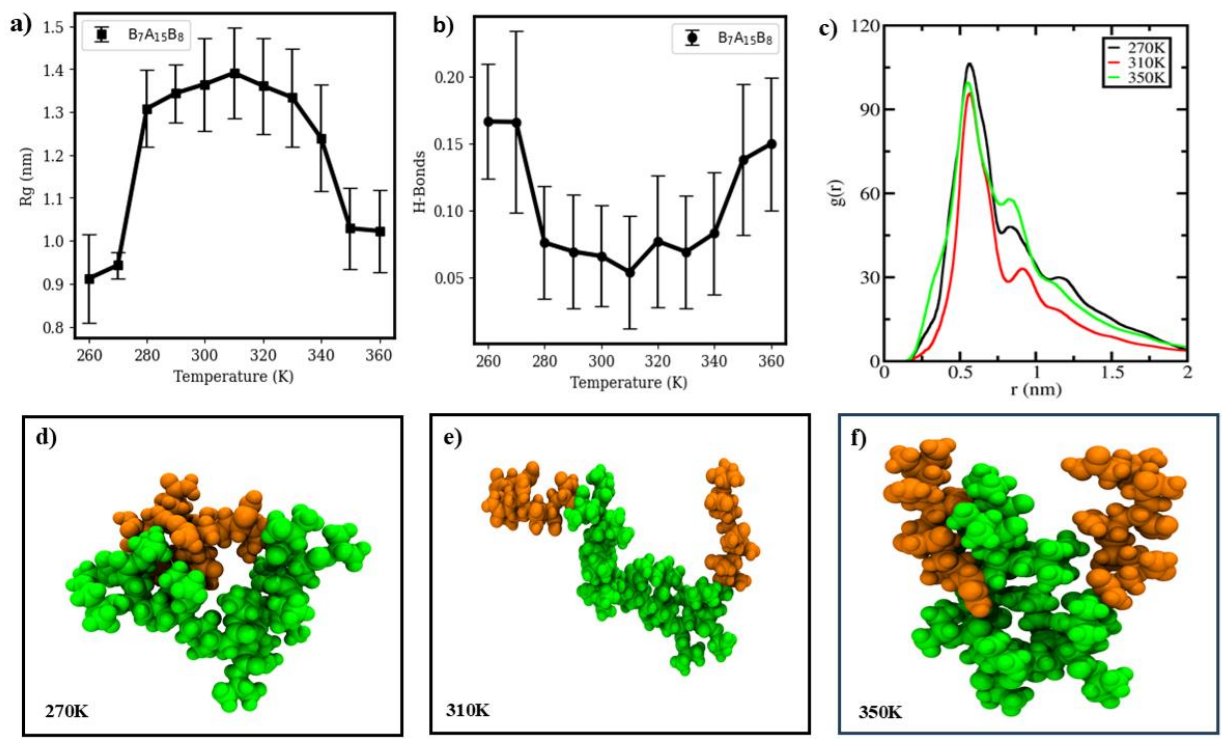

*Figure 3: Phase behaviour of $B_7A_{15}B_8$. The Rg is plotted as a function of temperature in (a). The number of polymer-polymer hydrogen bonds, normalized by the total number of monomer in a chain, is shown as a function of temperature in (b). The N-N RDF is plotted in (c). Three MD snapshots of the system corresponding to T= 270K, 310K and 350K are shown in (d), (e) and (f), respectively. Water moleucles are deleted for better visual inspections.*

We now turn our attention to the remaining copolymers, which either exhibit no phase transition or display both LCST and UCST behaviour within the temperature range considered in this study. These systems represent more complex thermoresponsive profiles, likely arising from a delicate balance between intra- and intermolecular interactions that govern solubility across a broader temperature range. First, we report the phase behaviour of $B_7A_{15}B_8$ in Figure 5. Interestingly, this particular copolymer undergoes two transitions. The $R_g$ of the polymer increases sharply as the temperature rises from 270 K to 280 K. This suggests a UCST-type crossover. The $R_g$ does not change significantly from 280K to 310K. Beyond 310K, the $R_g$ drops sharply, indicating an LCST transition. The number of monomer-monomer hydrogen bonds as a function of temperature is shown in Figure 5b. It indicates two crossover points corresponding to the LCST and UCST of the system. We also calculate the pair correlation function for the nitrogen atoms (N–N) of the polymer chain. As shown in Figure 5c, the first peak height decreases while moving from 270K to 300K. However, the peak height increases again as we move from 310K to 350K. This is consistent with the two transitions seen in the $R_g$-$T$ data. Interestingly, the phase transition disappears as we reverse the sequence, as shown in Figure 6 for $A_7B_{15}A_8$. For $A_7B_{15}A_8$, the $R_g$ does not change noticeably for the temperature range (260-360K). The total number of monomer-monomer hydrogen bonds also does not change significantly in this temperature window. The N-N pair correlation function remains essentially unchanged in this window. This highlights the profound impact that the special arrangement of chemical moieties has on its solution phase behavior. Finally, we consider a random sequence - $A_1B_1A_3B_2A_5B_4A_3B_5A_2B_3A_1$, whose phase behaviour is reported in Figure 7. This particular sequence shows two transitions. As the temperature increases, the $R_g$ drops sharply at 250 K. It again increases significantly beyond 340 K. The total number of monomer-monomer hydrogen bonds as a function of temperature also suggests two transitions (Figure 7b). As the temperature rises from 240K to 290K, there is an increase in hydrogen bonds due to the transition of polymer chains from a coil to a globular structure. Subsequently, the increase in temperature from 310K to 360K results in a decrease in hydrogen bonds as the polymer transitions from a globular to a coil structure. Similar to previous cases, we plot the N-N pair correlation function in Figure 7c, which correlates well with the $R_g$ trend with temperature.

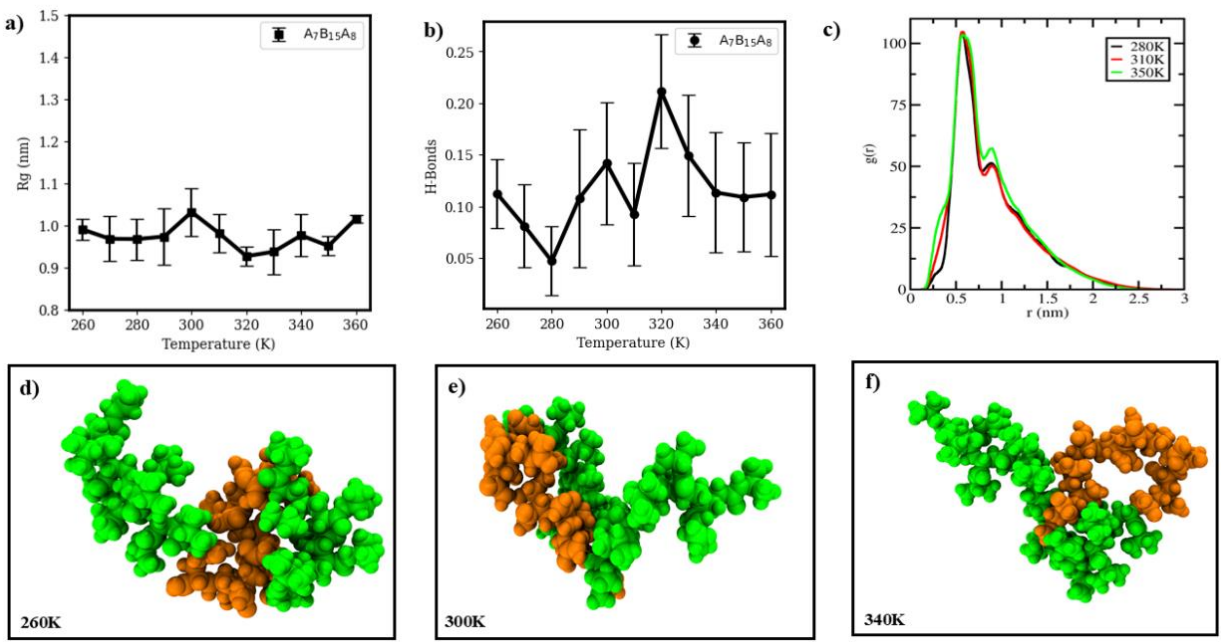

*Figure 6: Phase behaviour of $A_7B_{15}A_8$. The Rg is plotted as a function of temperature in (a). The number of polymer-polymer hydrogen bonds, normalized by the total number of monomers in a chain, is shown as a function of temperature in (b). The N-N RDF is plotted in (c). Three MD snapshots of the system corresponding to T= 260K, 300K and 340K are shown in (d), (e) and (f), respectively.*

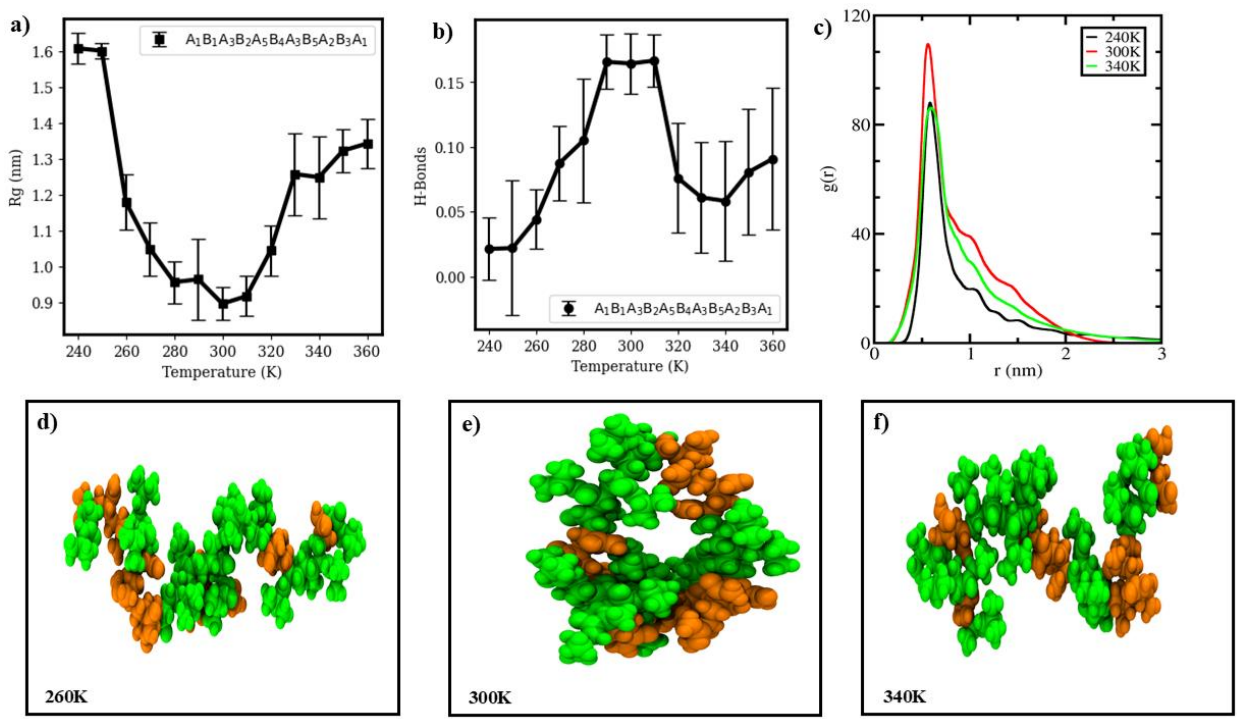

*Figure 7: Phase behaviour of $A_1B_1A_3B_2A_5B_4A_3B_5A_2B_3A_1$. The Rg is plotted as a function of temperature in (a). The number of polymer-polymer hydrogen bonds, normalized by the numbrer of monomers in a chain, is shown as a function of temperature in (b). The N-N RDF is plotted in (c). Three MD snapshots of the system corresponding to T= 260K, 300K and 340K are shown in (d), (e) and (f), respectively.*

## IV. Conclusions

The solution phase behavior of polymers is critically important for practical applications, and is of significant interest in both fundamental research. Polymers adopt conformations (coil, globule, etc.) that are sensitive to solvent quality, temperature, concentration, and molecular architecture. In this study, we investigate the sequence-dependent phase behaviour of PNIPAM-co-PAM copolymers in water using molecular dynamics simulations. Simulations are performed for the infinitely dilute condition at 1bar pressure for a temperature ranging from 260K to 360 K. The polymer is modelled using the OPLS force field, and we use TIP3P model of water for this study. The analysis focuses on the influence of monomer sequence distribution on intermolecular interactions and its connection to the polymer solubility behaviour. Our results clearly demonstrate that monomer sequence plays a pivotal role in modulating the intermolecular interaction and, therefore, thermoresponsive properties of these polymers. Specifically, most of the periodic sequences exhibited clear LCST-type phase transitions marked by a significant alteration in hydrogen bonding and a variability in LCST. We identified two nonperiodic sequences that show both LCST and UCST. Moreover, one sequence does not show any phase change in the explored temperature range. These findings highlight the importance of sequence engineering as a design parameter to tailor the phase behaviour of thermoresponsive polymers for various applications in smart materials, drug delivery, and temperature-sensitive devices.

**Acknowledgements.**

This work is made possible by financial support from the SERB, DST, Government of India through a core research grant (CRG/2022/006926).

**References.**